\documentclass[twocolumn,aps,prl,groupedaddress,nofootinbib,showpacs]{revtex4}
\usepackage{graphicx}
\usepackage{amsmath}
\usepackage{amssymb}
\usepackage{multirow}
\usepackage{bm}
\usepackage{color}
\usepackage{hyperref}

\usepackage{relsize}
\RequirePackage{xspace}
\usepackage{dcolumn}
\usepackage{bm}

\def\jpsi{{J/\psi}}
\def\psip{{\psi^\prime}}

\def\sa{{\bigl.^1\hspace{-1mm}S^{[8]}_0}}
\def\sb{{\bigl.^3\hspace{-1mm}S^{[8]}_1}}
\def\pj{{\bigl.^3\hspace{-1mm}P^{[8]}_J}}
\def\mo{\mathcal{O}}

\def\mopa{{\langle\mathcal{O}^\jpsi(\bigl.^1\hspace{-1mm}S_0^{[8]})\rangle}}
\def\mopb{{\langle\mathcal{O}^\jpsi(\bigl.^3\hspace{-1mm}S_1^{[8]})\rangle}}
\def\mopc{{\langle\mathcal{O}^\jpsi(\bigl.^3\hspace{-1mm}P_0^{[8]})\rangle}}
\def\mopsi{{\langle\mathcal{O}^{\jpsi}_n\rangle}}

\def\moppb{{\langle\mathcal{O}^\psip(\bigl.^3\hspace{-1mm}S_1^{[8]})\rangle}}
\def\moppc{{\langle\mathcal{O}^\psip(\bigl.^3\hspace{-1mm}P_0^{[8]})\rangle}}

\def\dssa{\hat{d\s}[\sa]}
\def\dssb{\hat{d\s}[\sb]}
\def\dspj{\hat{d\s}[\pj]}

\def\Majpsi{M_{0,r_0}^{\jpsi}}
\def\Mbjpsi{M_{1,r_1}^{\jpsi}}

\def\Mbpsip{M_{1,r_1}^{\psip}}
\def\Majpsip{M_{0,r_0}^{\jpsi(\psip)}}
\def\Mbjpsip{M_{1,r_1}^{\jpsi(\psip)}}

\def\MoH{{\langle\mathcal{O}^H\rangle}}
\def\MaH{M_{0,r_0}^{H}}
\def\MbH{M_{1,r_1}^{H}}
\def\ptcut{p_T^{cut}}
\def\be{\begin{equation}}
\def\ee{\end{equation}}
\def\bea{\begin{eqnarray}}
\def\eea{\end{eqnarray}}
\def\NO{\nonumber}
\def\mev{\mathrm{~MeV}}
\def\gev{\mathrm{~GeV}}
\def\tev{\mathrm{~TeV}}

\def\dfrac{\displaystyle\frac}

\def\a{\alpha}

\def\s{\sigma}

\def\muL{\mu_{\Lambda}}
\def\msbar{\overline{\mathrm{MS}}}

\begin{document}


\title{\mbox{}\\[10pt]
$\bm{\jpsi(\psip)}$ production at the Tevatron and LHC at
$\bm{\mo(\a_s^4v^4)}$ in nonrelativistic QCD}

\author{Yan-Qing Ma$~^{(a)}$, Kai Wang$~^{(a)}$, and Kuang-Ta Chao$~^{(a,b)}$}
\affiliation{ {\footnotesize (a)~Department of Physics and State Key
Laboratory of Nuclear Physics and Technology, Peking University,
 Beijing 100871, China}\\
{\footnotesize (b)~Center for High Energy Physics, Peking
University, Beijing 100871, China}}




\begin{abstract}
We present a complete evaluation for $\jpsi(\psip)$ prompt
production at the Tevatron and LHC at next-to-leading order in
nonrelativistic QCD, including color-singlet, color-octet, and
higher charmonia feeddown contributions.  The short-distance
coefficients of $\pj$ at next-to-leading order are found to be
larger than leading order by more than an order of magnitude but
with a minus sign at high transverse momentum $p_T$.  Two new linear
combinations of color-octet matrix elements are obtained from the
CDF data, and used to predict $\jpsi$ production at the LHC, which
agrees with the CMS data.  The possibility of $\sa$ dominance and
the $\jpsi$ polarization puzzle are also discussed.
\end{abstract}
\pacs{12.38.Bx, 13.60.Le, 14.40.Pq}

\maketitle

Nearly 20 years ago, the CDF Collaboration found a surprisingly
large production rate of $\psip$ at high $p_T$\cite{Abe:1992ww}. To
solve the large discrepancy between data and theoretical
predictions, the color-octet(CO) mechanism \cite{Braaten:1994vv} was
proposed based on nonrelativistic QCD (NRQCD)
factorization\cite{Bodwin:1994jh}. With the CO mechanism, $Q\bar{Q}$
pairs can be produced at short distances in CO ($\sa$, $\sb$, $\pj$)
states and subsequently evolve into physical quarkonia by
nonperturbative emission of soft gluons. It can be verified that the
partonic differential cross sections at leading-order (LO) in $\a_s$
behave as $1/p_T^4$ for $\sb$, and $1/p_T^6$ for $\sa$ and $\pj$,
all of which decrease at high $p_T$ much slower than $1/p_T^8$ of
the color-singlet (CS) state. The CO mechanism could give a natural
explanation for the observed $p_T$ distributions and large
production rates of $\psip$ and $\jpsi$ \cite{Kramer:2001hh}.

However, the CO mechanism seems to encounter difficulties in
explaining the observed $\jpsi(\psip)$ polarizations. Dominated by
gluon fragmentation to $\sb$, the LO NRQCD predicts transverse
polarization for $\jpsi(\psip)$ at high $p_T$\cite{Kramer:2001hh}
whereas measurements at the Fermilab Tevatron give almost
unpolarized $\jpsi(\psip)$\cite{Affolder:2000nn}. To exploit the
underlying physics, several efforts have been made, either by
introducing new channels\cite{Artoisenet:2007xi} or by proposing
other mechanisms\cite{Haberzettl:2007kj}. It is a significant step
to work out the next-to-leading order (NLO) QCD correction for the
CS channel, which enhances the differential cross section by about 2
orders of magnitude at high $p_T$\cite{Campbell:2007ws}, and changes
the polarization from being transverse at LO into longitudinal at
NLO\cite{Gong:2008sn}. Although the CS NLO cross section still lies
far below the experimental data, it implies that, compared to the
$\alpha_s$ suppression, kinematic enhancement at high $p_T$ is more
important in the current issue. This observation is also supported
by our recent work\cite{Ma:2010vd} for $\chi_c$ production, where we
find the ratio of production rates of
$\s_{\chi_{c2}}/\s_{\chi_{c1}}$ can be dramatically altered by the
NLO contribution due to change of the $p_T$ distribution from
$1/p_T^6$ at LO to $1/p_T^4$ at NLO in the CS P-wave channels. So we
may conclude nothing definite until  all important channels in
$1/p_T$ expansion are presented. It means the CO channels $\sa$ and
$\pj$ should be considered at NLO, while the CS channel
$^3S^{[1]}_1$  at next-to-next-to-leading order (NNLO) in
$\alpha_s$. Among these corrections, the complete NNLO calculation
for CS is beyond the state of the art, and the NNLO$^\star$ method
is instead proposed\cite{Lansberg:2008gk}, in which only tree-level
diagrams at this order are considered and an infrared cutoff is
imposed to control soft and collinear divergences, and the
NNLO$^\star$ contributions are shown to be large. However, the only
$1/p_T^4$ leading contribution at NNLO in CS is given by gluon
fragmentation, which was found\cite{braaten} to be negligible
compared to the observed $\jpsi(\psip)$ production data. Other NNLO
contributions may give a $1/p_T^6$ term. In a complete NNLO
calculation with both real and virtual corrections, infrared and
collinear divergences are removed and these NNLO $1/p_T^6$
contributions should be smaller than the NLO $1/p_T^6$ contribution
due to $\a_s$ suppression. Therefore, to achieve a good description
for $\jpsi(\psip)$ production a complete NLO calculation including
both CS and CO seems to be necessary.

At present, NRQCD factorization formalism with the CO mechanism is
used to describe various processes in heavy quarkonium production
and decay. While $\jpsi$ production in two-photon collisions at CERN
LEP2\cite{Klasen:2001cu} and photoproduction at DESY
HERA\cite{Chang:2009uj} are shown to favor the presence of CO
contribution, the $\jpsi$ production at $B$ factories is described
well using NLO CS model and leaves little room for CO
contributions\cite{Ma:2008gq}.   In order to further test the CO
mechanism, it is necessary to study hadroproduction and extract CO
long distance matrix elements (LDMEs) at NLO.

In view of the importance, here we present a complete NLO
contribution to $\jpsi(\psip)$ production at the Tevatron and LHC,
including all important CS and CO channels. According to the NRQCD
factorization formalism, the inclusive cross section for direct
$\jpsi$ production in hadron-hadron collisions is expressed as
\bea \label{eq:NRQCD} &d\s[pp\rightarrow \jpsi+X]=\sum\limits_{n}\hat{d\s}[(c\bar{c})_n]\dfrac{\mopsi}{m_c^{2L_n}}\\
&=\sum\limits_{i,j,n}\int \mathrm{d}x_1\mathrm{d}x_2 G_{i/p}G_{j/p}
\times\hat{d\s}[i+j\rightarrow (c\bar{c})_n +X]\mopsi, \NO\eea where
$p$ is either a proton or an antiproton, the indices $i, j$ run over
all the partonic species, and $n$ denote the color, spin and angular
momentum ($L_n$) of the intermediate $c\bar{c}$ states, including
$^3S^{[1]}_1$, $\sa$, $\sb$ and $\pj$ in the present issue. Compared
with the $S$-wave channel obtained in
\cite{Campbell:2007ws,Gong:2008sn,Gong:2008ft}, the NLO treatment of
$\pj$ is much more complicated. Fortunately, using the same method
as in \cite{Ma:2010vd}, we are able to get a compact expression for
the virtual correction, which is both time-saving and numerically
stable in the final state phase space integration. For technical
details, we refer readers to Ref.\cite{Ma:2010vd}.


For numerical results, we choose the same parameters as in
\cite{Ma:2010vd} except that here we are restricted to $\sqrt{S} =
1.96 \tev$ and $|y_{\jpsi(\psip)}| < 0.6$ with the Tevatron, while
$\sqrt{S} = 7 \tev$ and $|y_{\jpsi(\psip)}| < 2.4$ with the LHC.

\begin{figure}
\includegraphics[width=7.5cm]{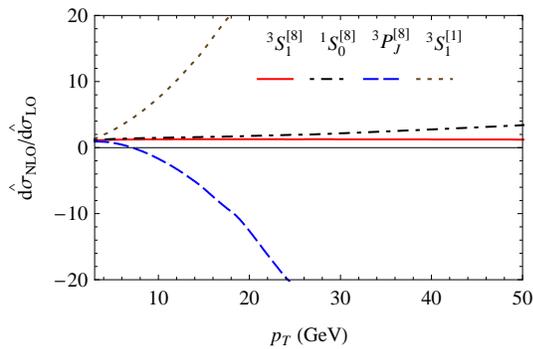}
\caption{\label{fig:kfactor} Dependence of $K$ factors (ratios of
NLO to LO short-distance coefficients $\hat{d\sigma}$) on $p_T$ in
$\jpsi(\psip)$ direct production at the Tevatron.}
\end{figure}

Let us first have a glance at the overall correction behaviors as
presented in Fig.~\ref{fig:kfactor}.
We find the K factor of short-distance coefficients $\hat{d\sigma}$
for $\pj$ channels (the sum over J=0,1,2 weighted with a factor of
2J+1 by spin symmetry in nonrelativistic limit) is large but
negative at high $p_T$. As explained in \cite{Ma:2010vd}, the
negative value mainly originated from using the $\msbar$ scheme when
choosing the renormalization scheme for S-wave spin-triplet NRQCD
LDMEs, and does not affect the physical result.
Another nontrivial phenomenon is that, differing from other
channels, the K factor of $\sb$ channel is almost independent of
$p_T$ and not larger than 1.3. This can be understood since the
$\a_s$ correction does not bring any new kinematically enhanced
contributions for the $\sb$ channel, and it implies the expansion in
$\a_s$ is under control once the leading $p_T$ (scaling as
$1/p_T^4$) channel is opened up.  We also note that K factors of all
other channels are just about 1 when $p_T \approx 3 \gev$, which can
be seen in Fig.~\ref{fig:kfactor}. All the large corrections can be
attributed to the enhancement in $1/p_T$ expansion.

\begin{figure}
\includegraphics[width=7.5cm]{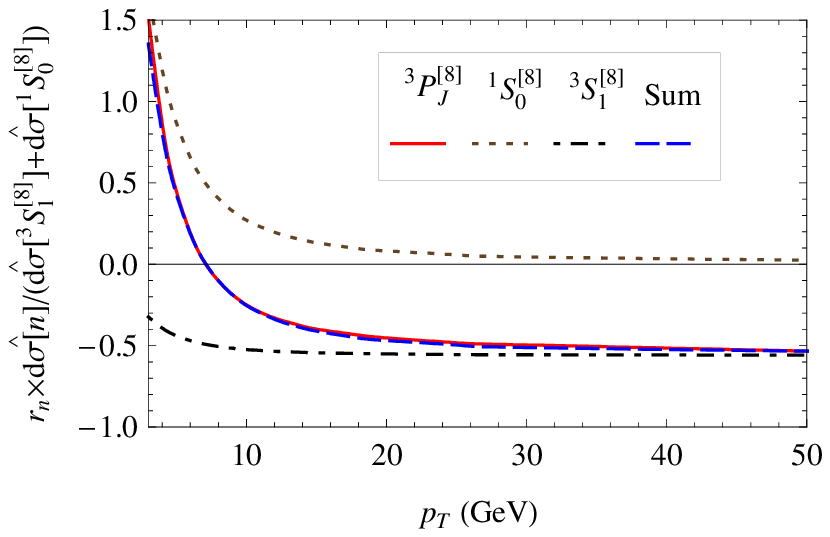}
\caption{\label{fig:decomposition} NLO short-distance coefficients
$\dspj$, $r_0\dssa$, $r_1\dssb$, and $Sum = r_0 \dssa + r_1 \dssb$
as functions of $p_T$ at the Tevatron, where $r_0$ = 3.9, $r_1$ =
-0.56 and each contribution is divided by $\dssa + \dssb$.}
\end{figure}

Since we find $\pj$ channels can give a $1/p_T^4$ term and have a
large K factor, the $\sb$ channel is no longer the unique source for
high $p_T$ contributions. In fact, for the short-distance
coefficients defined in Eq.~(\ref{eq:NRQCD}) the following
decomposition holds within an error of a few percent
\be \dspj = r_0~\dssa + r_1~\dssb, \ee
where we find $r_0 = 3.9$ and $r_1=-0.56$ for the Tevatron, and $r_0
= 4.1$ and $r_1=-0.56$ for the LHC. This decomposition in direct
$\jpsi(\psip)$ production at the Tevatron is shown in
Fig.\ref{fig:decomposition},
where each contribution is divided by $\dssa + \dssb$ to make it
easy to read. As a result, it is convenient to use two linearly
combined LDMEs
\bea \label{MEs} \Majpsi &=\mopa + \dfrac{r_0}{m_c^2}\mopc, \NO\\
\Mbjpsi &=\mopb + \dfrac{r_1}{m_c^2}\mopc, \eea when comparing
theoretical predictions with experimental data for production rates
at the Tevatron and LHC.

We note that, although both $\mopb$ and $\dspj$ depend on the
renormalization scheme and the factorization scale $\muL$, $\Mbjpsi$
does not. The reason is that the dependence of $\mopb$ is canceled
by that of $r_1$, which is originated from decomposing $\dspj$ at
high $p_T$ with all information for the dependence (here we ignore
the contribution of $^3S^{[1]}_1$, which decreases quickly at high
$p_T$ in LO). So $r_1$ should be viewed as
$r_1(\overline{\mathrm{MS}},\muL)$ but for simplicity we suppress
these variables in the expression.

\begin{figure}
\includegraphics[width=7.5cm]{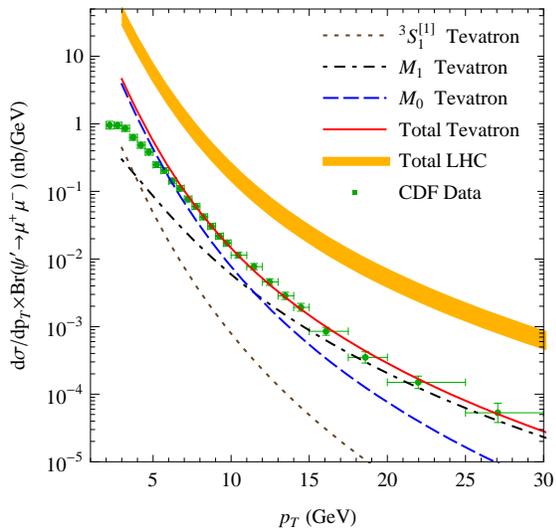}
\caption{\label{fig:psi2s} Transverse momentum distributions of
prompt $\psip$ production at the Tevatron and LHC. CDF data are
taken from Ref.\cite{Acosta:2004yw}. The yellow bands indicate the
uncertainty due to CO LDMEs.}
\end{figure}

\begin{figure}
\includegraphics[width=7.5cm]{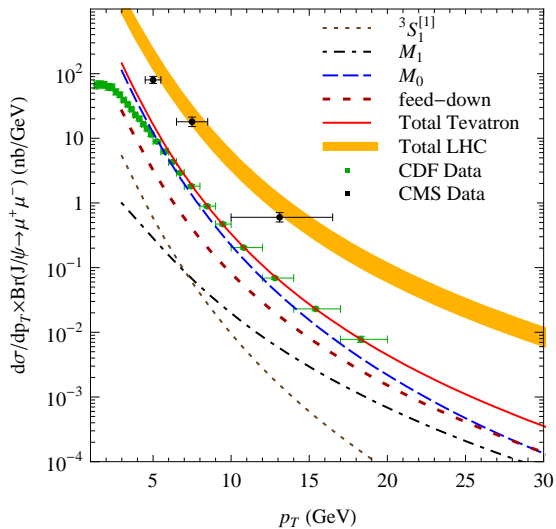}
\caption{\label{fig:jpsi} The same as Fig.~\ref{fig:psi2s} but for
$\jpsi$ production. The preliminary CMS data, taken from
Ref.\cite{cms}, are compared with the theoretical prediction. }
\end{figure}

\begin{table}
\begin {center}
\begin{tabular}{|c|c|c|c|c|c|}
 \hline
$\ptcut$&\multirow{2}{*}{\centering $H$}&$\MoH$& $\MbH$& $\MaH$&\multirow{2}{*}{\centering $\chi^2/d.o.f.$}\\
$\gev$& &$\gev^3$&$10^{-2}\gev^3$&$10^{-2}\gev^3$&~\\
\hline
\multirow{2}{*}{\centering $7$}&$\jpsi$&$1.16 $&$ 0.05\pm0.02$&$ 7.4\pm1.9$&$0.33$\\
~&$\psip$&$0.76 $&$ 0.12\pm0.03$&$ 2.0\pm0.6$&$$0.56\\
\hline
\multirow{2}{*}{\centering $5$}&$\jpsi$&$1.16 $&$ 0.16\pm0.05$&$ 5.2\pm1.3$&$3.5$\\
~&$\psip$&$0.76 $&$ 0.17\pm0.04$&$ 1.1\pm0.3$&$2.2$\\
\hline
\end{tabular}
\caption{Fitted color-octet LDMEs in $\jpsi(\psip)$ production with
chosen $\ptcut$. Here $r_0 = 3.9$, $r_1=-0.56$ are determined from
short-distance coefficient decomposition at the Tevatron. Errors are
due to renormalization and factorization scale dependence only.
Color-singlet ($^3S^{[1]}_1$) LDMEs $\MoH$ are estimated using a
potential model result\cite{Eichten:1995ch}.}
 \label{table2}
\end {center}
\vspace{-0.5cm}
\end{table}

By fitting the $p_T$ distributions of prompt $\psip$ and $\jpsi$
production measured at Tevatron\cite{Acosta:2004yw} in
Fig.~\ref{fig:psi2s} and Fig.~\ref{fig:jpsi}, the CO LDMEs are
determined as showing in Table \ref{table2}, while the CS LDMEs are
estimated using a potential model result of the wave functions at
the origin\cite{Eichten:1995ch}. In the fit we introduce a $\ptcut$
and only use experimental data for the region $p_T \geq \ptcut$. In
Figs.~\ref{fig:psi2s}and \ref{fig:jpsi} and the following analysis,
we prefer to use $\ptcut = 7 \gev$.

We find the ratio $R=\Mbjpsi / \Majpsi$ is determined to be as small
as 0.007. Based on this fit, we may conclude that the direct $\jpsi$
production could be dominated by the $\sa$ channel in the chosen
experimental $p_T$ region. To achieve this conclusion, we emphasize
the following points on the origination of the small $R$.

(1)~We find the fitted results are not good for data with $p_T < 7
\gev$, while the data for $p_T \geq 7 \gev$ can be fitted very well
using the determined LDMEs for both $\jpsi$ and $\psip$. We perform
a $\chi^2$ analysis for comparing theoretical fit with experimental
data with different $\ptcut$. Values of $\chi^2/d.o.f.$ decrease
rapidly as the cut increasing from $3 \gev$ to $7 \gev$, and
$\chi^2/d.o.f.$ becomes almost unchanged when $\ptcut$ is larger.
This may be understood as factorization and perturbation expansion
may not be reliable at low $p_T$. In Fig.~\ref{fig:jpsi} the
curvature of observed cross section is positive at large $p_T$ but
negative at small $p_T$, with a turning point at $p_T \approx
6\gev$. But the theoretical curvature is positive. This implies data
below $7 \gev$ may not be well explained in this work (even in
perturbative QCD) and needs further studying. Nevertheless, as an
alternative choice, we also give the fitted result for $\ptcut = 5
\gev$, for which $\Mbjpsi$ is increased by a factor of 3, while the
price paid is $\chi^2/d.o.f.$ increases from 0.33 to 3.5. The
results for both $\ptcut = 7 \gev$ and $\ptcut = 5 \gev$ are shown
in Table \ref{table2}.

(2)~Feed-down contributions from $\psip$ and $\chi_{cJ}$ to $\jpsi$
prompt production are properly considered. Because $m_{\psip}$ and
$m_{\chi_{cJ}}$ are larger than $m_{\jpsi}$ by only a few hundred
$\mev$,   $\jpsi$ is almost motionless in the higher charmonium rest
frame. So $p_T$ of $\jpsi$ can be expressed as $p_T\approx
p_T^{\prime}\times (m_{\jpsi}/{m_H})$,
where  $p_T^{\prime}$ and $m_H$ are the transverse momentum and mass
of the directly produced higher charmonium $H$.  LDMEs of $\psip$
are taken from Table \ref{table2}, while that of $\chi_{cJ}$ are
chosen with relatively smaller values from Ref.\cite{Ma:2010vd}.
From experimental data in Figs.~\ref{fig:psi2s} and \ref{fig:jpsi};
and Ref.\cite{Ma:2010vd}, we see that the prompt production $p_T$
distribution of $\jpsi$ is steeper than that of $\psip$ and
$\chi_{cJ}$. This implies that the subtraction of more feeddown
contributions will lead to a steeper $\jpsi$ direct production
distribution and hence a smaller $R$.

(3) Errors come from other sources. Varying renormalization and
factorization scales from $m_T/2$ to $2m_T$,  where
$m_T=\sqrt{4m_c^2+p_T^2}$ typically changes both $\Mbjpsi$ and
$\Majpsi$ by 30\% (Table \ref{table2}). However, the ratio $R$ is
almost independent of changing scales, because the dependence
between two LDMEs cancels each other. Varying the charm quark mass
$m_c$ can change the values of both LDMEs and $R$, and the
dependence of $R$ on $m_c$ is approximately $R \varpropto m_c^2$.
Thus choosing $m_c=1.5\pm0.1$ may cause an error of 20\% for $R$.

So, using the Tevatron  data of $\jpsi$ prompt production for $p_T
\geq 7 \gev$ or even $p_T \geq 5 \gev$, we find very small values
for $R$, or equivalently, $\Mbjpsi\ll \Majpsi$ (see Table
\ref{table2}). If we make a simple assumption that  the smallness of
$\Mbjpsi$ is not due to accidental cancellation between $\mopb$ and
$\mopc$,  we would have an order of magnitude estimate for the three
LDMEs \be \mopb \approx \mopc/m_c^2 \ll \mopa.\NO \ee This would
lead to a nontrivial result that $\jpsi$ direct production is
dominated by the $\sa$ channel, hence $\jpsi$ is mainly unpolarized,
which agrees with the polarization
measurement\cite{Affolder:2000nn}.

We have also compared our prediction for prompt $\jpsi$ production
with the CMS data in Fig.~\ref{fig:jpsi} and a good agreement is
achieved.

As for $\psip$, since the difference between two LDMEs is not as
large as that of $\jpsi$,  $\Mbpsip$ may be dominant at not too high
$p_T$; hence, $\psip$ may be transversely polarized in this region.
However, it should be noted that $\Mbpsip$ is always a combination
of $\moppb$ and $\moppc$ at NLO; thus, whether $\psip$ is
transversely polarized at high $p_T$ is unclear and needs further
studying.

In summary, we calculate $\jpsi(\psip)$ prompt production at the
Tevatron and LHC at $\mo(\a_s^4v^4)$, including all CS, CO, and
feeddown contributions. A large K factor of P-wave CO channels at
high $p_T$ results in two linearly combined LDMEs $\Majpsip$ and
$\Mbjpsip$, which can be extracted at NLO from the Tevatron data.
Because of the steep shape of experimental $\jpsi$ prompt production
data, we get a very small $\Mbjpsi$, which might indicate the
possibility that CO $\sa$ dominates $\jpsi$ direct production. If
this is the case, $\jpsi$ will be mainly unpolarized, which may
provide a possible solution to the long-standing $\jpsi$
polarization puzzle.


We thank C. Meng and Y.J. Zhang for helpful discussions, and B. Gong
and J.X. Wang for useful communications. This work was supported by
the National Natural Science Foundation of China (No.10721063, No.
11021092, No.11075002) and the Ministry of Science and Technology of
China (No.2009CB825200).

{\it{Note added.}} Soon after this work was submitted for
publication, a similar study appeared\cite{Butenschoen:2010rq}, and
for all color-singlet and octet channels in $J/\psi$ direct
hadroproduction their short-distance coefficients are consistent
with ours.




\begin{thebibliography}{}

\bibitem{Abe:1992ww}
  F.~Abe {\it et al.}  [CDF Collaboration],
  Phys.\ Rev.\ Lett.\  {\bf 69}, 3704 (1992).


\bibitem{Braaten:1994vv}
  E.~Braaten and S.~Fleming,
  Phys.\ Rev.\ Lett.\  {\bf 74}, 3327 (1995),
  [\href{http://arxiv.org/abs/hep-ph/9411365}{arXiv:hep-ph/9411365}].

\bibitem{Bodwin:1994jh}
  G.~T.~Bodwin, E.~Braaten and G.~P.~Lepage,
  Phys.\ Rev.\  D {\bf 51}, 1125 (1995),\  D {\bf 55}, 5853 (E)
  (1997),
  [\href{http://arxiv.org/abs/hep-ph/9407339}{arXiv:hep-ph/9407339}].

\bibitem{Kramer:2001hh}
 M.~Kr\"amer,
  Prog.\ Part.\ Nucl.\ Phys.\  {\bf 47}, 141 (2001),
  [\href{http://arxiv.org/abs/hep-ph/0106120}{arXiv:hep-ph/0106120}];
See also N.~Brambilla {\it et al.},
\href{http://arxiv.org/abs/hep-ph/0412158}{arXiv:hep-ph/0412158}.

\bibitem{Affolder:2000nn}
  A.~A.~Affolder {\it et al.}  [CDF Collaboration],
  Phys.\ Rev.\ Lett.\  {\bf 85}, 2886 (2000),
  [\href{http://arxiv.org/abs/hep-ex/0004027}{arXiv:hep-ex/0004027}];
  A.~Abulencia {\it et al.}  [CDF Collaboration],
  Phys.\ Rev.\ Lett.\  {\bf 99}, 132001 (2007),
  [\href{http://arxiv.org/abs/0704.0638}{arXiv:0704.0638}].

\bibitem{Artoisenet:2007xi}
  P.~Artoisenet, J.~P.~Lansberg and F.~Maltoni,
  Phys.\ Lett.\  B {\bf 653}, 60 (2007),
  [\href{http://arxiv.org/abs/hep-ph/0703129}{arXiv:hep-ph/0703129}];
  Z.~G.~He, R.~Li and J.~X.~Wang,
  Phys.\ Rev.\  D {\bf 79}, 094003 (2009),
  [\href{http://arxiv.org/abs/0904.2069}{arXiv:0904.2069}].

\bibitem{Haberzettl:2007kj}
  G.~C.~Nayak, J.~W.~Qiu and G.~Sterman,
  Phys.\ Lett.\  B {\bf 613}, 45 (2005),
  [\href{http://arxiv.org/abs/hep-ph/0501235}{arXiv:hep-ph/0501235}];
  Phys.\ Rev.\  D {\bf 72}, 114012 (2005),
  [\href{http://arxiv.org/abs/hep-ph/0509021}{arXiv:hep-ph/0509021}];
  H.~Haberzettl and J.~P.~Lansberg,
  Phys.\ Rev.\ Lett.\  {\bf 100}, 032006 (2008),
  [\href{http://arxiv.org/abs/0709.3471}{arXiv:0709.3471}].

\bibitem{Campbell:2007ws}
  J.~M.~Campbell, F.~Maltoni and F.~Tramontano,
  Phys.\ Rev.\ Lett.\  {\bf 98}, 252002 (2007),
  [\href{http://arxiv.org/abs/hep-ph/0703113}{arXiv:hep-ph/0703113}].

\bibitem{Gong:2008sn}
  B.~Gong and J.~X.~Wang,
  Phys.\ Rev.\ Lett.\  {\bf 100}, 232001 (2008),
 [\href{http://arxiv.org/abs/0802.3727}{arXiv:0802.3727}];
  Phys.\ Rev.\  D {\bf 77}, 054028 (2008),
 [\href{http://arxiv.org/abs/0805.2469}{arXiv:0805.2469}].

\bibitem{Ma:2010vd}
  Y.~Q.~Ma, K.~Wang and K.~T.~Chao,
  \href{http://arxiv.org/abs/1002.3987}{arXiv:1002.3987}.

\bibitem{Lansberg:2008gk}
P. Artoisenet et al., Phys. Rev. Lett. 101, 152001 (2008),
 [\href{http://arxiv.org/abs/0806.3282}{arXiv:0806.3282}];
  J.~P.~Lansberg,
  Eur.\ Phys.\ J.\  C {\bf 61}, 693 (2009),
  [\href{http://arxiv.org/abs/0811.4005}{arXiv:0811.4005}].

\bibitem{braaten}
  E.~Braaten, M.~A.~Doncheski, S.~Fleming and M.~L.~Mangano,
  Phys.\ Lett.\  B {\bf 333}, 548 (1994).

\bibitem{Klasen:2001cu}
  M.~Klasen, B.~A.~Kniehl, L.~N.~Mihaila and M.~Steinhauser,
  Phys.\ Rev.\ Lett.\  {\bf 89}, 032001 (2002),
  [\href{http://arxiv.org/abs/hep-ph/0112259}{arXiv:hep-ph/0112259}].

\bibitem{Chang:2009uj}
  P.~Artoisenet, J.~M.~Campbell, F.~Maltoni and F.~Tramontano,
  Phys.\ Rev.\ Lett.\  {\bf 102}, 142001 (2009),
  [\href{http://arxiv.org/abs/0901.4352}{arXiv:0901.4352}];
  C.~H.~Chang, R.~Li and J.~X.~Wang,
  Phys.\ Rev.\  D {\bf 80}, 034020 (2009),
  [\href{http://arxiv.org/abs/0901.4749}{arXiv:0901.4749}];
  M.~Butenschoen and B.~A.~Kniehl,
  Phys. Rev. Lett. 104, 072001 (2010),
  [\href{http://arxiv.org/abs/0909.2798}{arXiv:0909.2798}].

\bibitem{Ma:2008gq}
  Y.~Q.~Ma, Y.~J.~Zhang and K.~T.~Chao,
  Phys.\ Rev.\ Lett.\  {\bf 102}, 162002 (2009),
  [\href{http://arxiv.org/abs/0812.5106}{arXiv:0812.5106}];
  B.~Gong and J.~X.~Wang,
  Phys.\ Rev.\ Lett.\  {\bf 102}, 162003 (2009),
  [\href{http://arxiv.org/abs/0901.0117}{arXiv:0901.0117}];
Y.J.~Zhang, Y.Q.~Ma, K.~Wang, and K.T.~Chao, Phys. Rev. D81, 034015
(2010),
   [\href{http://arxiv.org/abs/0911.2166}{arXiv:0911.2166}];
  Y.~J.~Zhang and K.~T.~Chao,
  Phys.\ Rev.\ Lett.\  {\bf 98}, 092003 (2007),
  [\href{http://arxiv.org/abs/hep-ph/0611086}{arXiv:hep-ph/0611086}].

\bibitem{Gong:2008ft}
  B.~Gong, X.~Q.~Li and J.~X.~Wang,
  Phys.\ Lett.\  B {\bf 673}, 197 (2009),
  [\href{http://arxiv.org/abs/0805.4751}{arXiv:0805.4751}].

\bibitem{Acosta:2004yw}
  D.~E.~Acosta {\it et al.}  [CDF Collaboration],
  Phys.\ Rev.\  D {\bf 71}, 032001 (2005),
  [\href{http://arxiv.org/abs/hep-ex/0412071}{arXiv:hep-ex/0412071}];
  T.~Aaltonen {\it et al.}  [CDF Collaboration],
  Phys.\ Rev.\  D {\bf 80}, 031103 (2009),
  [\href{http://arxiv.org/abs/0905.1982}{arXiv:0905.1982}].

\bibitem{cms}
  N.~Leonardo, PoS {\bf ICHEP2010}, 207 (2010).

\bibitem{Eichten:1995ch}
  See the B-T model in E.~J.~Eichten and C.~Quigg,
  Phys.\ Rev.\  D {\bf 52}, 1726 (1995),
  [\href{http://arxiv.org/abs/hep-ph/9503356}{arXiv:hep-ph/9503356}].

\bibitem{Butenschoen:2010rq}
  M.~Butenschoen and B.~A.~Kniehl,
  Phys.\ Rev.\ Lett.\  {\bf 106}, 022003 (2011),
[\href{http://arxiv.org/abs/1009.5662}{arXiv:1009.5662}].



\end{thebibliography}

\end{document}